\documentclass{article}
\usepackage{frascatiphys,graphicx,subfigure}
\newcommand{\ket}{\,\rangle}
\newcommand{\bra}{\langle \,}

\def\m{\mu}
\def\n{\nu}

\begin{document}
\title{{\bf HADRONIC $\tau$ DECAYS INTO TWO AND THREE MESON MODES WITHIN RESONANCE CHIRAL THEORY}}
\author{ Pablo Roig\\
{\em IFIC (CSIC-Universitat de Val\`{e}ncia) 
 and Physik Department (TUM-Munchen)}}
\maketitle
\baselineskip=11.6pt
\begin{abstract}
We study two and three meson decays of the tau lepton within the framework of the Resonance Chiral Theory, that is based on the following properties of $QCD$: its chiral symmetry in the massless case, its large-$N_C$ limit, and the asymptotic behaviour it demands to the relevant form factors.\\
Most of the couplings in the Lagrangian are determined this way rendering the theory predictive. Our outcomes can be tested thanks to the combination of a very good experimental effort (current and forthcoming, at B- and tau-charm-factories) and the very accurate devoted Monte Carlo generators.
\end{abstract}
\baselineskip=14pt
\section{Hadronic decays of the $\tau$ lepton}
Our purpose is to provide a description of the semileptonic decays of the tau lepton that incorporates as many theoretical restrictions derived from the fundamental interaction, $QCD$ \cite{QCD}, as possible. This is a very convenient scenario to investigate the hadronization of $QCD$  because one fermionic current is purely leptonic and thus calculable unambiguously 
 so that we can concentrate our efforts on the other one, involving light quarks coupled to a $V-A$ current
.\\
The decay amplitude for the considered decays may be written as:
\begin{equation} \label{Mgraltau}
\mathcal{M}\,=\,-\frac{G_F}{\sqrt{2}}\,V_{\mathrm{ud/us}}\,\overline{u}_{\nu_\tau}\gamma^\mu(1-\gamma_5)\,u_\tau \mathcal{H}_\mu\,,
\end{equation}
where the strong interacting part is encoded in the hadronic vector, $\mathcal{H}_\mu$:
\begin{equation} \label{Hmugral}
\mathcal{H}_\mu = \bra \left\lbrace  P(p_i)\right\rbrace_{i=1}^n |\left( \mathcal{V}_\mu - \mathcal{A}_\mu\right)  e^{i\mathcal{L}_{QCD}}|0\ket\,.
\end{equation}
Symmetries let us decompose $\mathcal{H}_\mu$ depending on the number of final-state pseudoscalar ($P$) mesons, $n$.\\
One meson tau decays can be predicted in terms of the measured processes $(\pi/K)^-\to\mu^-\nu_\mu$,  since the matrix elements are related. This provides a precise test of charged current universality \cite{ToniTau}. On the other side, it cannot tell anything new on hadronization.\\
The two-pion tau decay is conventionally parameterized -in the isospin limit, that we always assume- just in terms of the vector ($J^P=1^-$) form factor of the pion, $F_\pi(s)$:
\begin{equation}
 \bra \pi^-\pi^0|\bar{d}\gamma^\mu u| 0\ket \equiv\bra \pi^-\pi^0|V^\mu \mathrm{e}^{i \mathcal{L}_{QCD}}| 0\ket \equiv\sqrt{2}F_\pi(s)(p_{\pi^-}-p_{\pi^0})^\mu\,,
\end{equation}
where $s\equiv(p_{\pi^-}+p_{\pi^0})^2$.\\
Because $SU(3)$ is broken appreciably by the difference between $m_s$ and $(m_u+m_d)/2$, two form factors are needed to describe the decays involving one pion and one kaon:
\begin{eqnarray}
 \bra \pi^-(p')\bar{K}^0(p)|V^\mu \mathrm{e}^{i \mathcal{L}_{QCD}}| 0\ket & \equiv & \left(\frac{Q^\mu Q^\nu}{Q^2}-g^{\m\n}\right)(p-p')_\nu F_+^{K\pi}(Q^2)\,\nonumber\\
& & -\,\frac{m_K^2-m_\pi^2}{Q^2}Q^\m F_0^{K\pi}(Q^2)\,,
\end{eqnarray}
where $Q^\m\equiv (p+p')^\m$. $F_+^{K\pi}(Q^2)$ carries quantum numbers $1^-$, while $F_0^{K\pi}(Q^2)$ is the pseudoscalar form factor ($0^-$). Other two meson decays can be treated similarly and one should take advantage of the fact that chiral symmetry relates some of their matrix elements.\\
 For three mesons in the final state, the most general decomposition reads:
\begin{eqnarray} \label{Hmu3m}
\mathcal{H}_\mu = V_{1\mu} F_1^A(Q^2,s_1,s_2) + V_{2\mu} F_2^A(Q^2,s_1,s_2) +\nonumber\\
 Q_\mu F_3^A(Q^2,s_1,s_2) + i \,V_{3\mu} F_4^V(Q^2,s_1,s_2)\,,
\end{eqnarray}
and
\begin{eqnarray} \label{VmuQmu}
V_{1\mu} &  = & \left( g_{\mu\nu} - \frac{Q_{\mu}Q_{\nu}}{Q^2}\right) \,
(p_2 - p_1)^{\nu} ,\,\,\,\,V_{2\mu} = \left( g_{\mu\nu} - \frac{Q_{\mu}Q_{\nu}}{Q^2}\right) \,
(p_3 - p_1)^{\nu}\,,\nonumber\\
V_{3\mu} & = & \varepsilon_{\mu\nu\varrho\sigma}\,p_1^\nu\, \,p_2^\varrho\, \,p_3^{\sigma} , \,\,\,\,Q_\mu  =  (p_1\,+\,p_2\,+\,p_3)_\mu \,,\,s_i = (Q-p_i)^2\,.
\end{eqnarray}
$F_i$, $i=1,2,3$, correspond to the axial-vector current ($\mathcal{A}_\m$) while $F_4$ drives the vector current ($\mathcal{V}_\m$). The form factors $F_1$ and $F_2$ have a transverse structure in the total hadron momenta, $Q^\mu$, and drive a $J^P=1^+$ transition. The pseudoscalar form factor, $F_3$, vanishes as $m_P^2/Q^2$ and, accordingly, gives a tiny contribution. Higher-multiplicity modes can be described proceeding similarly \cite{Fischer:1979fh}. This is as far as we can go without model assumptions, that is, it is not yet known how to obtain the $F_i$ from $QCD$. However, one can derive some of their properties from the underlying theory, as we will explain in the following.
\section{Theoretical framework: Resonance Chiral Theory}
We use a phenomenological Lagrangian \cite{Weinberg:1978kz} written in terms of the relevant degrees of freedom that become active through the energy interval spanned by hadronic tau decays. The chiral symmetry of massless $QCD$ determines \cite{ChPT} the chiral invariant operators that can be written including the lightest mesons in the spectrum, the pseudoscalar ones belonging to the pion multiplet. It was carefully checked \cite{Colangelo:1996hs} that -as one expects- Chiral Perturbation Theory, $\chi PT$, can only describe a little very-low-energy part of semileptonic tau decays.\\
Then one may attempt to extend the range of applicability of $\chi PT$ to higher energies while keeping its predictions for the form factors at low momentum: this is the purpose of Resonance Chiral Theory, $R \chi T$, \cite{RChT} that includes the light-flavoured resonances as explicit fields in the action.\\
At $LO$ in the $N_C\to\infty$ limit of $QCD$ \cite{Nc} one has as infinite tower of stable mesons that experience local effective interactions at tree level. We depart from this picture in two ways:
\begin{itemize}
 \item We incorporate the widths of the resonances worked out consistently within $R\chi T$ \cite{GomezDumm:2000fz}.
 \item We attempt a description including the least possible number of resonance fields reducing -ideally- to the single resonance approximation, $SRA$ \cite{SRA}.
\end{itemize}
We take vector meson dominance into account when writing our Lagrangian. Thus, it will consist of terms accounting for the following interactions ($a_\mu$, $v_\mu$ stand for the axial(-vector) currents and $A$ and $V$ for the axial(-vector) resonances):
\begin{itemize}
\item Those in $\chi PT$ at $LO$ and $\chi PT$-like: $a_\mu P$, $a_\mu PPP$, $PPPP$ with even-intrinsic parity; and $v_\mu PPP$ in the odd-intrinsic parity sector.
\item Those relevant in $R\chi T$ -$NLO$ $\chi PT$ operators are not included to avoid double counting, since they are recovered and their couplings are saturated upon integration of the resonance contributions \cite{RChT}-. They include: $a_\mu VP$, $v_\mu V$, $a_\mu A$, $VPP$ and $AVP$ in the even-intrinsic parity sector and $v_\mu VP$, $VPPP$ and $VVP$ in the odd-parity one.
\end{itemize}
The explicit form of the operators and the naming for the couplings can be read from \cite{RChT},  \cite{VVP}, \cite{tau3pi}, \cite{VAP}, \cite{proc} and \cite{paper}.\\
The $R\chi T$ just determined by symmetries does not share the $UV$ $QCD$ behaviour yet. For this, and for our purposes, we need to impose appropriate Brodsky-Lepage conditions \cite{BL} on the relevant form factors. Explicit computation and these short-distance restrictions reduce appreciably the number of independent couplings entering the amplitudes which enables us to end up with a useful -that is, predictive- theory.
\section{Phenomenology}
Our framework describes pretty well the two-meson decays of the $\tau$, as shown in the $\pi\pi$ \cite{pipi} and $K\pi$ \cite{Kpi} cases. Two-meson modes including $\eta$ can be worked analogously. The data in these modes are so precise \footnote{See the references quoted in the articles cited through the section.} that although the $SRA$ describes the gross features of the data, one needs to include the first excitations of the $V$ resonances to achieve an accurate description. Although $F_+^{K\pi}(Q^2)$ is much more important than $F_0^{K\pi}(Q^2)$, one needs an appropriate pseudoscalar form factor to fit well the data, specially close to threshold \cite{Kpi}.\\
The three meson modes are much more involved. However, a good description of the data has been achieved through a careful study \cite{tau3pi, paper} taking into account all theory constrains and experimental data on the $3\pi$ and $KK\pi$ channels. We predict the $KK\pi$ spectral function and conclude that the vector current contribution cannot be neglected in these modes. We will study the other three meson modes along the same lines. In particular, our study of the $K\pi\pi$ channels might help improve the simultaneous extraction of $m_s$ and $V_{us}$ \cite{Gamiz:2004ar}. Our expressions for the vector and axial-vector widths and the hadronic matrix elements have been implemented successfully in the TAUOLA library \cite{Olga}. This way, the experimental comunity will have as its disposal a way of analysing hadronic decays of the tau that includes as much as possible information from the fundamental theory. We also plan to study $e^+e^-\to PPP$ at low energies what can eventually be used by \cite{PHOKHARA}.\\
\section{Acknowledgements}
I congratulate the local organizing committee for this first edition of the Young Researchers Workshop as well as for the XIV LNF SPRING SCHOOL BRUNO TOUSCHEK held in parallel. I am grateful to Jorge Portol\'es for a careful revision and useful suggestions on the draft. I acknowledge useful discussions with Olga Shekhovtsova. This work has been supported in part by the EU MRTN-CT-2006-035482 (FLAVIAnet) and by the DFG cluster of excellence 'Origin and structure of the Universe'
.

\end{document}